\documentstyle[sprocl]{article}
\def\ra{\rightarrow}

\begin{document}
\begin{flushright}
UH-511-913-98 \\
August 1998
\end{flushright}
\vspace{.25in}
\title{CP VIOLATION IN HYPERON DECAYS\footnote{Presented at the Workshop on CP Violation,
Adelaide, July 3-8, 1998.}}
\author{SANDIP PAKVASA}
\address{Department of Physics \& Astronomy,\\
University of Hawaii \\
Honolulu, HI  96822 USA}
%
%

\maketitle
\abstract{The theory and phenomenology of CP violation in hyperon
decays is summarized.}


\section{Introduction}

In 1958 Okubo\cite{okubo} observed that CP violation allows hyperons and
antihyperons to have different branching ratios into conjugate channels
even though their total rates must be equal by CPT.  Later this paper
inspired Sakharov\cite{sakha} to his famous work on cosmological baryon-antibaryon
asymmetry.  Pais\cite{pais} extended Okubo's proposal to asymmetry parameters in
$\Lambda$ and $\bar{\Lambda}$ decays. Only now, some 40 years later are
these proposals about to be tested in the laboratory\cite{pais}.

The reason for the current interest is the need to find CP violation in
places other than just $K_L-K_S$ mixing.  Only a number of different
observations of CP violation in different channels will help us pin down
the source and nature of CP violation in or beyond the standard model.
>From this point of view hyperon decay is one more piece in addition to
the B system, the D system, $\epsilon'/\epsilon$ etc.

\section{Phenomenology of Hyperon Decays.}

I summarize here the salient features of the phenomenology of
non-leptonic hyperon decays \cite{rosen}.  Leaving out $\Omega^-$ decays, there  are
seven decay modes $\Lambda \ra N \pi, \ \Sigma^\pm \ra N \pi$ and $\Xi \ra \Lambda \pi$.  The
effective matrix element can be written as 
\begin{equation}
i \ \bar{u}_{\bar{p}} (a + b \gamma_5) u_\Lambda \ \phi
\end{equation}
for the mode $\Lambda \ra p + \pi^-$, where a and b are complex in general.
The corresponding element for $\bar{\Lambda} \ra \bar{p} + \pi^+$ is
then:
\begin{equation}
i \ \bar{v}_{\bar{p}} (-a^* + b^* \gamma_5) v_{\bar{\Lambda}} \phi^+
\end{equation}
It is convenient to express the observables in terms of S and  P and
write the matrix element as
\begin{equation}
S + P \ {\bf \sigma}.\hat{\bf q}
\end{equation}
where {\bf q} is the proton momentum in the $\Lambda$ rest frame
and S and P are:
\begin{eqnarray}
S &=& a \sqrt{ \left\{ (m_\Lambda + m_p)^2 - m^2_\pi \right\}
\over 
16 \pi \ m^2_{\Lambda}} \nonumber \\
P &=& b \sqrt {\left\{ (m_\Lambda - m_p)^2 - m^2_\pi \right\}
\over 16 \pi \ m^2_{\Lambda}}
\end{eqnarray}
In the $\Lambda$ rest-frame, the decay distribution is given by:
\begin{eqnarray}
\frac{d \Gamma}{d \Omega} &=&
\frac{\Gamma}{8 \pi} %
\{ 
[1 + \alpha < {\bf \sigma}_\Lambda > . \hat{\bf \sigma} ] \nonumber \\
&+& < {\bf \sigma}_p >. [( \alpha + < {\bf \sigma}_\Lambda > . 
\hat{\bf q}) \hat{\bf q}+ \ \ \beta < {\bf \sigma}_\Lambda > \times \hat{\bf q} \nonumber \\
&+& \gamma ( \hat{\bf q} \times ( < {\bf \sigma}_\Lambda > \times
 \hat{\bf q})] 
 \}
\end{eqnarray}  
where $\Gamma$ is the decay rate and is given by:
\begin{equation}
\Gamma = 2 \mid {\bf q} \mid \left\{ \mid S \mid^2 + \mid P \mid^2 \right \}
\end{equation}
$\alpha, \beta$ and $\gamma$ are given by
\begin{eqnarray}
\alpha &=& 2Re (S^*P ) \over \left \{ \mid S \mid^2 + \mid P \mid^2 \right
\}, \nonumber \\
\beta &=& 2 Im (SP^*) \over \left \{ \mid S \mid^2 + \mid P \mid^2
\right \}
\nonumber \\
\gamma &=&  \left \{ \mid S \mid^2 - \mid P \mid^2 \right \} \over 
\left \{ \mid S \mid^2 + \mid P \mid^2 \right \}
\end{eqnarray}
For a polarized $\Lambda$, the up-down asymmetry of the final proton is
given by  $\alpha ( \alpha$ is also the longitudinal polarization of the
proton for an unpolarized $\Lambda)$.  $\beta$ and $\gamma$ are
components of the transverse polarization of proton \cite{lee}.

The observed properties of the hyperon decays can be summarised as:  (i)
the $\Delta I = 1/2$ dominance i.e. the $\Delta I = 3/2$ amplitudes are about
5\% of the $\Delta I = 1/2$ amplitudes; (ii) the asymmetry parameter
$\alpha$ is large for $\Lambda$ decays, $\Xi$ decays and $\Sigma^+ \ra p
\pi^0$ and is near zero for $\Sigma^\pm \ra n \pi^\pm$; and (iii) the
Sugawara-Lee triangle sum rule $\sqrt{3}A ( \Sigma^+ \ra p  \pi^0) - A 
(\Lambda \ra p \pi^-) = 2A (\Xi \ra \Lambda \pi^-)$ is satisfied to a level
of 5\%.

\section{CP Violating Observables}

Let a particle P decays into several final states $f_1, f_2$ etc.  The
amplitude for P $\ra f_1$ is in general:
\begin{equation}
A= A_1 e^{i \delta 1} \ \ + A_2 \ e^{i \delta 2}
\end{equation}
where 1 and 2 are strong interaction eigenstates and $\delta_i$ are
corresponding final state phases.  Then the amplitude for
$\bar{P} \ra \bar{f}_1$ is
\begin{equation}
\bar{A} = A^*_1 e^{i \delta 1} \ \ + A^*_2 \ e^{i \delta 2}
\end{equation}
If $\mid A_1 \mid >> \mid A_2 \mid$, then the rate asymmetry
$\Delta ( = (\Gamma - \bar{\Gamma}) / (\Gamma + \bar{\Gamma}) )$ is
given by:
\begin{equation}
\Delta \approx -2 \mid A_2/A_1 \mid sin ( \phi_1 - \phi_2) sin(\delta_1- 
\delta_2)
\end{equation}
where $A_i = \mid A_i \mid e^{i \phi_i}$.  Hence, to get a non-zero rate
asymmetry, one must have (i) at least two channels in the final state,
(ii) CPV weak phases must be different in the two channels, and (iii) 
\underline{unequal} final state scattering phase shifts in the two
channels. These conditions were delineated by Brown et al\cite{brown}.  
A similar calculation of the asymmetry of $\alpha$ shows that
for a single isospin channel dominance,
\begin{equation}
A= (\alpha + \bar{\alpha})/ (\alpha - \bar{\alpha}) = 2 \tan ( \delta_s -
\delta_p) \ \tan (\phi_s - \phi_p)
\end{equation}
In this case the two channels are orbital angular momentum $0$ and $1$;
and even a single isospin mode such as $\Xi^- \ra \Lambda \pi^-$ can
exhibit a non-zero A.

To define the complete set of CP violating observables, consider the
example of the decay modes $\Lambda \ra p \pi^-$ and
$\bar{\Lambda} \ra \bar{p} \pi^+$.  The amplitudes are:
\begin{eqnarray}
S&=& - \sqrt{2 \over 3} S_1 e^{i ( \delta_1 + \phi_1^s)} +
    \frac{1} {\sqrt{3}} S_3 e^{i ( \delta_3 + \phi^s_3)} \nonumber \\
P&=& - \sqrt{2 \over 3} P_1 e^{i ( \delta_{11} + \phi_1^p)} +
    \frac{1}{\sqrt{3}} P_3 e^{i ( \delta_3 + \phi^p_3)} 
\end{eqnarray}
where $S_i, P_i$ are real, $i$ refers to the final state isospin (i=2I)
and $\phi_i$ are the CPV phases.  With the knowledge that $S_3/S_1$,
$P_3/P_1 <<$ 1 ;  one can write\cite{donog}
\begin{eqnarray}
\Delta_\Lambda &=& \frac{(\Gamma-\overline{\Gamma})}{(\Gamma + \overline{\Gamma})}
\cong \sqrt{2} \ (S_3/S_1) sin ( \delta_3 - \delta_1) sin
(\phi_3^s - \phi_1^s) \nonumber \\
A_\Lambda &=& \frac{(\alpha +\overline{\alpha})} {(\alpha - \overline{\alpha})}
\cong -\tan (\delta_{11} - \delta_1) \tan (\phi_1^p - \phi_1^s)
\nonumber \\
B_\Lambda &=& \frac{(\beta +\overline{\beta})}{(\beta - \overline{\beta})}
\cong  cot  (\delta_{11} - \delta_1) \tan (\phi_1^p - \phi_1^s)
\end{eqnarray}
For $\pi$N final states, the phase shifts at $E_{c.m.} = m_\Lambda$ are
known and are: $\delta_1 = 6^0, \ \delta_3 = -3.8^0, \ \delta_{11} = 1.1^0$
and $\delta_{31} = -0.7^0$.  The CPV phases $\phi_i$ have to be provided
by theory.

Similar expressions can be written for other hyperon decays.  For
example, for $\Lambda \ra n \pi^0$, $\Delta$ is $- 1/2 \Delta_\Lambda$
and $A$ and $B$ are identical to $A_\Lambda$ and $B_\Lambda$.  For
$\Xi^- \ra \Lambda \pi^-$ ( and $\Xi^0 \ra \Lambda \pi^0)$ the 
asymmetries are:
\begin{eqnarray}
\Delta_\Xi & = & 0    \nonumber \\
A_\Xi & = &-tan (\delta_{21} - \delta_2) \tan (\phi^p - \phi^s)
\nonumber \\
B_\Xi & = & cot (\delta_{21} - \delta_2) \tan (\phi^p- \phi^s)
\end{eqnarray}
where $\delta_{21}$ and $\delta_2$ are the $p$ and $s$-wave $\Lambda
\pi$ phase shifts at $m_\Xi$ respectively.  Somewhat more
complicated expressions can be written for $\Sigma$ decays but I will
not exhibit them since there is no near-term experimental interest.

\section{Calculating CP Phases}

In standard model description of the non-leptonic hyperon decays, the
effective $\Delta S = 1$ Hamiltonian is
\begin{equation}
H_{eff} = \frac{G_F}{\sqrt{2}} \ U_{ud}^* \ U_{us} \sum^{12}_{i=1} c_i
(\mu) \ O_i ( \mu)
\end{equation}
after the short distance QCD corrections  (LLO + NLLO) where
$c_i = z_i + y_i \tau ( \tau = -U_{td} \ U^*_{ts} / U_{ud} \ U_{us})$,
and $\mu \sim 0(1$ GeV). 
For CP violation, the most important operator is:
\begin{equation}
O_6 = \bar{d} \ \lambda_i\gamma_\mu (1+ \gamma_5) s 
\bar{q} \lambda_i \gamma_\mu (1 -\gamma_5)q
\end{equation}
and $y_6 \approx -0.1$ at $\mu \sim 1 GeV, \ m_t \sim 175$ GeV and 
$\Lambda_{QCD} \sim 200-300$ MeV.  
To estimate the CP phases in the Eq. (12), one assumes that the real part of 
the amplitude has been correctly obtained.  The major 
uncertainty is the hadronic matrix elements and the fact that the 
simultaneous fit of S and P waves leaves a factor of 2 ambiguity
\cite{donog1}.  
In the SM, 
with the Kobayasi-Maskawa phase convention there is no CPV in $\Delta I 
= 3/2$ amplitudes; and for $\Lambda$ decays $\phi_3 = 0$.  Evaluating the 
matrix elements in the standard way and with the current knowledge of the 
K-M matrix one finds for the decays\cite{donog2} $\Lambda \ra p \pi^-$ and $\Xi^-
 \ra \Lambda \pi^-$:
\begin{eqnarray}
& &\phi^s_\Lambda - \phi^p_\Lambda \cong 3.5.10^{-4}  \nonumber \\
& &\phi_\Xi^s - \phi^p_\Xi \cong - 1.4.10^{-4}
\end{eqnarray}
With the $N \pi$  phase shifts known to be 
\begin{equation}
\delta_s - \delta_p \cong 7^0
\end{equation}
one finds for the asymmetry $A_\Lambda$ in the standard model a value of 
about $-4.10^{-5}$.   For the $\Xi \ra \Lambda \pi^-$ decay mode the phase 
shifts are not known experimentally and have to be determined 
theoretically.  There are calculations from 1965 \cite{martin} which gave large values for 
$\delta_s - \delta_p \sim 20^0$; however, all recent calculations based on chiral 
perturbation theory, heavy baryon approximation etc. agree that $\delta_s - 
\delta_p$ lies between $1^0$ and $3^0$ \cite{lu}.  In this case the asymmetry 
$A_\Xi$ is expected to be $\sim - (0.2$ to $0.7) 10^{-5}$.

An experimental measurement of the phase shifts $\delta_s - \delta_p$ in 
$\Lambda \pi$ system will put the predictions for $A_\Xi$ on a firmer basis.  
There is an old proposal due to Pais and Treiman \cite{pais1} to measure $\Lambda \pi$ 
phase shifts in $\Xi \ra \Lambda \pi ev$, but this does not seem
practical in near future.  Another technique, more feasible, it to measure $\beta$ and 
$\alpha$ to high precision in $\Xi$ and $\overline{\Xi}$ decays.  Then the 
combination.
\begin{equation}
( \beta - \bar{\beta}) / (\alpha - \bar{\alpha}) = \tan \ (\delta_s -
\delta_p)
\end{equation}
can be used to extract $\delta_s -\delta_p$.  To the extend CP phases are 
negligible one can also use the approximate relation:
\begin{equation}
\beta/\alpha \approx \tan (\delta_s -\delta_p)
\end{equation}

\section{Beyond Standard Model:}

Can new physics scenarios in which the source of CP violation is not K-M 
matrix yield large enhancements of these asymmetries?   We consider some 
classes of models where these asymmetries can be estimated reliably \cite{donog}.

First there is the class of models which are effectively super-weak \cite{some}.  
Examples are models in which the K-M matrix is real and the observed CP 
violation is due to exchange of heavier particles; heavy scalars with FCNC, 
heavy quarks etc.  In all such models direct CP violation is negligible and 
unobservable and so all asymmetries in hyperon decays are essentially zero.

In the three Higgs doublet model with flavor conservation imposed, the 
charged Higgs exchange tend to give large effects in direct CP violation as 
well as large neutron electric dipole moment \cite{wein}.

We consider two versions of left-right symmetric models:  (i) Manifest Left-
Right symmetric model without $W_L - W_R$ mixing \cite{mohap} and (ii) with $W_L -
W_R$ mixing \cite{chang}.  In (i) $U_{KM}^L =$ real and $U_{KM}^R$ complex 
with arbitrary phases but angles given by $U_{KM}^L$.  Then one gets 
the ``isoconjugate'' version in which
\begin{equation}
H_{eff} = \frac{G_F \ U_{us}} {\sqrt{2}}
\left [ J^\dagger_{\mu L} \ J_{\mu L} + \ \eta e^{i \beta} J^\dagger_{\mu R} \ J_{\mu
R}
\right ]
\end{equation}
where $\eta = m^2_{WL} /m^2_{WR}$ and $\beta$ is the relevant CPV phase.  Then 
$H_{p.c.}$ and $H_{p.v.}$ have overall phases $(1 + i \eta \beta)$ and
$(1-i \eta  \beta)$ respectively.  To account for the observed CPV in
K-decay, $\eta \beta$ 
has to be of order $4.5.10^{-5}$.  In this model, $\epsilon'/\epsilon = 0$ 
and there are no rate asymmetries in hyperon decays but the asymmetries
 A and B are not zero and e.g. A goes as 
$2 \eta \beta \sin (\delta_s - \delta_p)$.  In the class of models where
$W_L - W_R$ mixing is allowed, the asymmetries can be enhanced.

\begin{table}
\caption{Expectations for Hyperon CPV Asymetries.}
\begin{center}
\begin{tabular}{|lccccc|} \hline
                   &   \mbox{SM}   & \mbox{2-Higgs}    &   \mbox{FCNC}  &  L-R-S  &  L-R-S  \\ 
                   &               &    	       &  \mbox{Superweak}  & (1)   &  (2) \\ 
$\Delta_\Lambda$   &  $10^{-6}$    &   $10^{-5}$       &        0
                                                                            &   0    &  0  \\
$A_\Lambda$   &  $ -4.10^{-5}$    &   $-2.10^{-5}$  &      0  & $-10^5$ &   $6.10^{-4}$  \\
$B_\Lambda$   &$10^{-4}$             &   $2.10{-3}$     &  0    &
                   $7.10^{-4}$  & -  \\
$\Delta_\Xi$   & 0  & 0   &  0   &  0  &  0  \\
$A_\Xi$   &  $-4.10^{-6}$    &   $-3.10^{-4}$  & 0    & $2.10^{-5}$     & $10^{-4}$  \\
$B_\Xi$   &  $10^{-3}$    &   $ 4.10^{-3}$  & 0    & $3.10^{-4}$     &   -
                   \\ \hline
\end{tabular}
\end{center}
\end{table}

\section{Experiments}

There have been several proposals to detect  hyperon decay asymmetries in
$\bar{p} p \ra \bar{\Lambda} \Lambda, \ \bar{p}{p} \ra \overline{\Xi} \Xi$ and in 
$e^+e^- \ra J/\psi \ra \Lambda \overline{\Lambda}$ but none of these 
were approved \cite{hamman}.  The only approved and on-going experiment is E871 at 
Fermilab.  In this experiment $\Xi^-$ and $\overline{\Xi}^+$ are produced 
and the angular distribution of $\Xi^-  \ra \Lambda \pi^- \ra p \pi^- \pi^-$ and 
$\overline{\Xi}^+$ compared.  This effectively measures $A_\Lambda + 
A_\Xi^-$ and will be discussed by Luk \cite{luk}.

\section{$\epsilon'/\epsilon$ and Hyperon Decay Asymmetries.}

One might think that measurement of $\epsilon'/\epsilon$ already will check 
direct CPV in the $\Delta S= 1$ channel.  Why is it worthwhile measuring 
another $\Delta S=1$ process like hyperon decay?  The point is that there 
are important differences and the two are not at all identical.  A 
comparison is given below in Table 2
\begin{table}
\caption{Comparison of $\epsilon'/\epsilon$ with hyperons.}
\begin{center}
\begin{tabular}{|lll|} \hline
     & $\epsilon'/\epsilon$      &  CPV Asymmetry  \\
     &                           &  $A_\Lambda \ + \ A_\Xi$  \\
Matrix element    &  Large cancellation    &     \\
at large $m_t$    &  between e.m. and strong 	   &   No such cancellation   \\
               &     penguin &         \\
   &           &                 \\
SM range       &  $3.10^{-3}$  to $10^{-4}$      &   $A= -5.10^{-5}$ \\
               &                                 &(to within factor of
2) \\
   &   &  \\
Amplitude      &   Pure P.V.			&  Probes Interference \\
               &  (No information               &   between P.V. and P.C. \\
               &  on P.C. amplitudes)           &  Sensitive to new
physics \\
               &                               & in P.C. channel. \\ \hline
\end{tabular}
\end{center}
\end{table}
which makes it clear that $\epsilon'/\epsilon$ and hyperon decay 
asymmetries are different and complimentary.  The hyperon decay 
measurements are as important and significant as $\epsilon'/\epsilon$.

\section*{Conclusion}

The searches for direct CPV are being pursued in many channels.  $K \ra 2 
\pi, \Lambda \ra N {\pi}$, B decays and D decays.  Any observation of 
a signal would be the first outside $K^0 \overline{K}^0$ mixing and would 
rule out a large class of superweak models. Eventually we will be able to 
confirm or demolish the Standard Kobayashi-Maskawa description of CP 
violation.

\section*{Acknowledgment}

I acknowledge useful discussions with Alakabha Datta, Gene Golowich,
Xiao-Gang He, Kam-Biu Luk, Mahiko Suzuki, and German Valencia recently and with John Donoghue long
ago.  
The hospitality of Tony Thomas and 
colleagues and the stimulating atmosphere of the workshop was 
outstanding.  This work is supported in part by USDOE under Grant \#DE-
FG-03-94ER40833.


\section*{References}

\begin{thebibliography}{99}

\bibitem{okubo} S. Okubo, {\it Phys. Rev.} 109, 984 (1958).

\bibitem{sakha}A.D. Sakharov, Zh. EK. Teor. Fiz. 5, 32 (1967) (English
translation, {\it JETP Letters} 5, 24 (1967).

\bibitem{pais}A. Pais, {\it Phys. Rev. Lett.} 3, 242 (1959).

\bibitem{rosen}S. P. Rosen and S. Pakvasa, {\it Advances in Particle
Physics}, Ed. R.E. Marshak and R.L. Cool, Wiley-Interscience, NY 1968,
p. 473; S. Pakvasa and S.P. Rosen, ``{\it The Past Decade in Particle
Theory}'', Ed. E.C.G. Sudarshan and Y. Ne'eman, Gordon and Breach ,
1973, p. 437.

\bibitem{lee}T. D. Lee and C. N. Yang, {\it Phys. Rev. 108} 1645
(1957); R. Gatto, {\it Nucl. Phys. 5}, 183 (1958).
 
\bibitem{brown}T. Brown, S. Pakvasa and S. F. Tuan, {\it
Phys. Rev. Lett.} 51, 1823 (1983), see also O. E. Overseth and
S. Pakvasa, {\it Phys. Rev.} 184, 1663 (1969); L-L. Chau and 
H. Y. Cheng, Phys. Lett. B131, 202 (1983); D. Chang and L. Wolfenstein,
{\it Intense Medium Energy Sources of Strangeness}, edited by
T. Goldman, H.E. Haber, and H.F.-W. Sadrozinski, AIP Conf. Proc. No. 102
(AIP, New York, 1983), p. 73;
C. Kounnas, A. B. Lahanas and P. Pavlopoulos, {\it Phys. Lett.} B127,
381 (1983).

\bibitem{donog}J. F. Donoghue and S. Pakvasa, {\it Phys. Rev. Lett.} 55,
162 (1985).

\bibitem{donog1} J. F. Donoghue, E. Golowich, and B. Holstein,
{\it Phys. Rep.} 131, 319 (1986); {\it Dynamics of the Standard Model},
Cambdridge Univ. Press, 1992.

\bibitem{donog2} J. F. Donoghue, X-G. He and S. Pakvasa, {\it Phys. Rev.} D39,
833 (1986); M. J. Iqbal and G. Miller, {\it Phys. Rev.} D41, 2817
(1990); X-G. He, H. Steger and G. Valencia, {\it Phys. Lett.} B272, 411
(1991); N.G. Despande, X-G. He and S. Pakvasa, {\it Phys. Lett.} B326,
307 (1994); X-G. He and G. Valencia, {\it Phys. Rev.} D52, 5257 (1995).

\bibitem{martin}B. Martin, {\it Phys. Rev.} 138, 1136 (1965); R. Nath
and A. Kumar, {\it Nuov. Cim.} 36, 669 (1965).

\bibitem{lu}M. Lu, M. Wise and M. Savage, {\it Phys. Lett.} B337,
A. Datta and S. Pakvasa, {\it Phys. Lett.} B344, 340 (1995); A. Kamal,
hep-ph/9801349;
A. Datta, P.J. O'Donnell and S. Pakvasa, hep-ph/9806374.
 
\bibitem{pais1} A. Pais and S.B. Treiman, {\it Phys. Rev.} 178, 2365
(1969).

\bibitem{some} Some examples can be found in the following references:
R.N. Mohapatra, J. C. Pati and L. Wolfenstein, {\it Phys.  Rev.} D11,
3319 (1975);
T. Brown, N. Despande, S. Pakvasa and H. Sugawara, {\it Phys. Lett. B}
141, 95 (1984); J. M. Soares and L. Wolfenstein, {\it Phys. Rev.} D46,
256 (1992); P. Frampton and S.L. Glashow, {\it Phys. Rev.} D55, 1691
(1997); H. Georgi and S.L. Glashow, hep-ph/9807399.

\bibitem{wein} S. Weinberg, {\it Phys. Rev. Lett.} 37, 657 (1976).
 
\bibitem{mohap} R. N. Mohapatra and J. C. Pati, {\it Phys. Rev.} D11, 566
(1975).

\bibitem{chang} D. Chang, X-G. He and S. Pakvasa, {\it Phys. Rev. Lett.} 74,
3927 (1995).

\bibitem{hamman} N. Hamman et al. CERN/SPSPLC 92-19, SPSLC/M49;
S. Y. Hsueh and P. Rapidis, FNAL Proposal.

\bibitem{luk}K-B. Luk, these proceedings.
\end{thebibliography}
\end{document}